\documentclass[11pt,twoside]{article}
\usepackage{./asp2014}

\usepackage{mathrsfs}

\aspSuppressVolSlug
\resetcounters

\begin{document}

\title{Non-isothermal effects on Be disks}
\author{Rodrigo G. Vieira,$^1$ Alex C. Carciofi,$^1$ and Jon E. Bjorkman$^2$
\affil{$^1$Instituto de Astronomia, Geof\'isica e Ci\^encias Atmosf\'ericas, Universidade de S\~ao Paulo, Rua do Mat\~ao, 1226, Cidade Universit\'aria, 05508-090, S\~ao Paulo, Brazil; \email{rodrigo.vieira@iag.usp.br}, \email{carciofi@usp.br}}
\affil{$^2$Ritter Observatory, Department of Physics \& Astronomy, University of Toledo, Toledo, OH 43606, USA; \email{jon.bjorkman@utoledo.edu}}
}

\paperauthor{Rodrigo G. Vieira}{rodrigo.vieira@iag.usp.br}{}{Instituto de Astronomia, Geof\'isica e Ci\^encias Atmosf\'ericas, Universidade de S\~ao Paulo, Universidade de S\~ao Paulo}{Departamento de Astronomia}{S\~ao Paulo}{SP}{05508-900}{Brazil}
\paperauthor{Alex C. Carciofi}{carciofi@usp.br}{}{Instituto de Astronomia, Geof\'isica e Ci\^encias Atmosf\'ericas, Universidade de S\~ao Paulo, Universidade de S\~ao Paulo}{Departamento de Astronomia}{S\~ao Paulo}{SP}{05508-900}{Brazil}
\paperauthor{Jon E. Bjorkman}{jon.bjorkman@utoledo.edu}{}{Ritter Observatory, University of Toledo}{Department of Physics \& Astronomy}{Toledo}{OH}{43606}{USA}

\begin{abstract}
In the last decade, the {\it viscous decretion disk} model has emerged as the new paradigm for Be star disks. In this contribution, we propose a simple analytical model to estimate the continuum infrared excess arising from these circumstellar disks, in the light of the currently accepted scenario. We demonstrate that the disk can be satisfactorily described by a two component system: an inner optically thick region, which we call the {\it pseudo-photosphere}, and a diffuse outer part. In particular, a direct connexion between the disk brightness profile and the thermal structure is derived, and then confronted to realistic numerical simulations. This result quantifies how the non-isothermality of the disk ultimately affects both infrared measured fluxes and visibilities.
\end{abstract}

\section{Introduction}
The presence of gaseous circumstellar material around Be stars leads to continuum emission in addition to the stellar one. Such excess can be orders of magnitude larger than the stellar flux at longer wavelengths, and depends on both stellar and disk physical parameters.\\

The idea of describing the circumstellar material of a Be star by a disk-shaped geometry dates from the seminal work of \cite{struve1931}. Many years later, \cite{gehrz1974} realized that the material around Be stars is essentially gaseous, and the infrared (IR) excess arises from free-free emission in the disk. Based on this information, \cite{wright1975} developed the basic theory to describe the IR excess arising from an expanding envelope. \cite{lamers1984} and \cite{waters1986} have extended this model, by introducing a new formalism that includes a bound-free opacity contribution and an opening angle to the former outflowing envelope. \cite{cote1987} were able to interpret IRAS observations with this model, and then derive the disk density structure for a large sample of Be stars.\\

In the last decade, the viscous decretion disk model (VDD; \citealt{lee1991}) has emerged as the new paradigm for Be star disks. The VDD model makes relatively simple and falsifiable predictions for the disk structure, and it remains succesful to the many tests it was subjected in the last years (for a complete review, see \citealt{rivi2013}). Besides, the database of IR observations now available is much bigger than the IRAS database available for \cite{cote1987}. The new observational information makes possible the study of large samples with better data quality and unprecedented statistical completeness. Therefore, it is time to revisit the model approach provided by \cite{lamers1984}, in the light of the new status of the study of Be star disks.\\

In this contribution, we introduce a simple, yet quite realistic, analytical model to describe the IR continuum emission of a gaseous disk. As a first application of the model, we focus on the question of how non-isothermal effects modify the disk brightness profile, and consequently influences the emitted flux.

\section{Realistic radiative transfer results}
\label{compare}

In order to study the brightness profile of the disk, we have computed realistic models using {\ttfamily HDUST} \citep{carciofi2006}. {\ttfamily HDUST} is a three-dimensional Monte Carlo radiative transfer code, capable of simultaneously calculating the non-LTE hydrogen level populations, ionization fraction and electron temperature at each disk position. For our models, we have adopted a parametrized density structure compatible with the VDD model. It varies radially as a power-law, and is vertically described by a gaussian profile. The models were calculated for three disk base densities ($8.4\times 10^{-12}$, $2.8\times 10^{-11}$ and $8.4\times 10^{-11}$ ${\rm g/cm^{3}}$), while the other disk parameters were held fixed (density radial exponent $n=3.5$, scale height radial exponent $\beta=1.5$, pole-on disk orientation). We have also adopted stellar parameters of a $B1$ spectral subtype star \citep{allen1977}, and computed fluxes and images at a wavelength range from $1\,{\rm {\mu}m}$ to $50\,{\rm {\mu}m}$.\\

\begin{figure}[!t]
\begin{center}
\includegraphics[width=1 \columnwidth,angle=0]{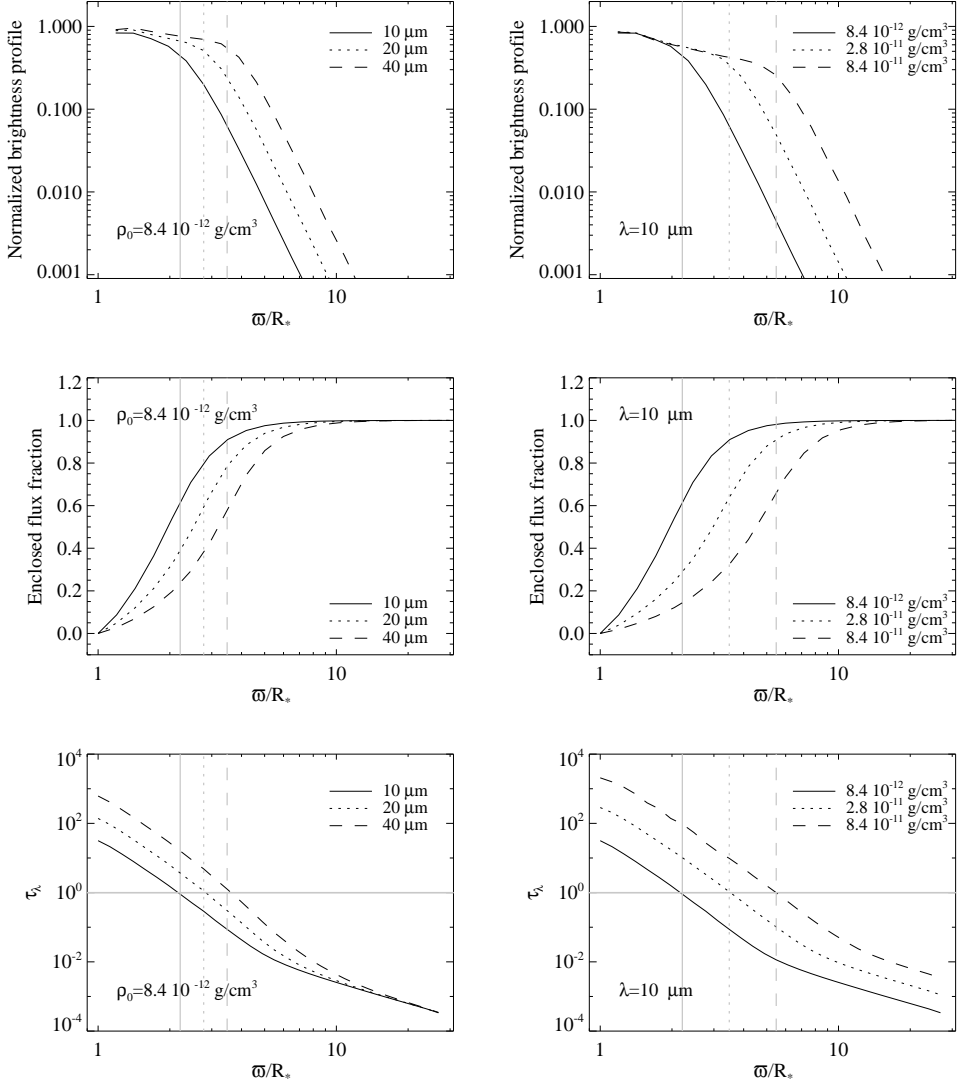}
\caption{Results from the models computed with {\ttfamily HDUST}. {\it Upper panels}: brightness profiles; {\it middle panels}: enclosed flux fractions; and {\it bottom panels}: vertical optical depth as a function of radius. The vertical lines indicate the position where $\tau_{\lambda}=1$ for each model. See text for more details of these models.}
\label{loci}
\end{center}
\end{figure}

Figure~\ref{loci} shows the radial brightness profiles and the respective enclosed flux fractions extracted from the images computed with {\ttfamily HDUST}. \cite{carciofi2011} and \cite{rivi2013} discuss the formation loci of different observables in terms of the enclosed flux fraction. For a pole-on disk, it can be defined as

\begin{equation} \label{def_loci}
\mathscr{L}_{\lambda}(R)=\frac{1}{F_{\lambda}^{disk}\,d^2}\,\int_{1}^{R} I_{\lambda}(\varpi) \, 2\pi\,\varpi \, d\varpi,
\end{equation}
where $I_{\lambda}$ is the disk specific intensity, $\varpi$ is the cylindrical radial coordinate, $R$ is an arbitrary position in the disk, and

\begin{equation}
F_{\lambda}^{disk}=\frac{1}{d^2}\,\int_{1}^{R_{disk}} I_{\lambda}(\varpi) \, 2\pi\,\varpi \, d\varpi.
\end{equation}
where $R_{disk}$ is the disk size. Conversely, we have that

\begin{equation}
I_{\lambda}(R)=\frac{F_{\lambda}^{disk}\,d^2}{2\,\pi\,R}\,\frac{\partial\mathscr{L}_{\lambda}(R)}{\partial R}.
\end{equation}

We can see from these relations that both specific intensity and enclosed flux fractions express the same physical information, the brightness distribution over the disk. The former can be seen as a differential expression, while the latter represents an integral description of this information. In this sense, they may therefore be considered equivalent definitions.\\

The numerical results suggest a dual brightness regime for the disk, with the inner part presenting a much flatter slope than the outer part. Typical values of a power-law equivalent slope are shown in Fig.~\ref{loci}. Besides, the transition between these two regions is marked by a ``knee'', at a distance from the star that clearly increases with both wavelength and base density. The simulations also indicate that the optical depth in the vertical direction is very close to the unity at this position.\\

Motivated by these results, we propose a new model assumes explicitly two emission regimes for the disk, an inner optically thick part and an outer optically thin part, separated by an effective radius ($\overline{R}$). We will interchangeably use the terms effective radius and {\it pseudo-photosphere} (disk region inside $\overline{R}$), as an analogy to the stellar photosphere.\\

In the next section, we derive the expressions that ensue from the above assumptions. A more complete model description and validation will be further presented by \citeauthor{vieira2015} (in preparation).

\section{The pseudo-photosphere model}
\label{pphot}

In order to describe the brightness profile of the disk at a given wavelength, we assume the effective radius to be the position in the disk where the optical depth in the vertical direction is equal to one. Mathematically, we have

\begin{equation}\label{ref_def}
\tau_{\lambda}(\overline{R})\equiv 1.
\end{equation}

To solve this equation for $\overline{R}$, we have first to derive $\tau_{\lambda}$ as a function of radius. For that purpose, we adopt the opacity expression given by \cite{lamers1984} in the Rayleigh-Jeans regime

\begin{equation}
\kappa_{\lambda} = 0.018\, T_d^{-3/2}\,\gamma\, \overline{z^2}\, \left(\frac{\rho}{\mu\,m_H}\right)^2\, (\lambda/c)^2\, [g(\lambda, T_d)+b(\lambda, T_d)],
\end{equation}
where $T_d$ is the disk temperature, $\gamma$ is the ionization fraction, $\overline{z^2}$ is the  squared atomic charge, $\mu$ the mean molecular density, $g$ and $b$ are respectively the free-free and bound-free gaunt factors. Furthermore, we adopt a parametric prescription for the disk density physically motivaded by the VDD model ({\it e.g.}, \citealt{carciofi2011})

\begin{equation}
\rho(\varpi,z)=\rho_0\,\left(\frac{\varpi}{R_{\star}}\right)^{-n}\exp\left(\frac{z^2}{2\,H^2} \right),
\end{equation}
where

\begin{equation}
H(\varpi)=H_0\,\left(\frac{\varpi}{R_{\star}} \right)^{\beta},
\end{equation}
and
\begin{equation}
H_0 = \frac{c_s}{v_{\phi}}\,R_{\star} \simeq \sqrt{\frac{k\,T_d(\varpi)}{\mu m_H} \frac{R_{\star}^3}{G M_{\star}}}.
\end{equation}

At this point, it is worth noticing that $H_0$ itself is {\it also a function of the position in the disk}, since it depends on the thermal structure $T_d$. The optical depth in the vertical direction can be written as

\begin{equation}
\tau_{\lambda} = \int_{-\infty}^{\infty} \kappa_{\lambda} dz = \tau_0\,\frac{T_{\star}}{T_d(\varpi)}\, \left(\frac{\varpi}{R_{\star}} \right)^{-2n+\beta}.
\end{equation}
where we define

\begin{equation}
\tau_0=\frac{0.018}{T_{\star}}\,\gamma\, \overline{z^2}\, \left(\frac{\rho_0}{\mu\,m_H}\right)^2\,\left(\frac{\pi\,k}{\mu m_H} \frac{R_{\star}^3}{G M_{\star}}\right)^{1/2}\,(\lambda/c)^2\, [g(\lambda, T_d)+b(\lambda, T_d)].
\end{equation}

The solution of Eq.~\ref{ref_def} depends on the previous knowledge of the disk thermal structure. Such structure can be estimated from the radiative equilibrium at each position of the disk. Except for some idealized situations, performing these calculations represents a very complicated task, that usually requires the implementation of realistic radiative transfer codes.

Nevertheless, the study of simplified situations can still provide useful informations about the disk properties. For example, if we adopt the simplification of a power-law temperature radial profile of the form

\begin{equation}
T_d(\varpi)=f\,T_{\star}\,\left(\frac{\varpi}{R_{\star}}\right)^{-s}
\end{equation}
where $0<f<1$, the solution for Eq.~\ref{ref_def} is then given by:

\begin{equation}\label{ref}
\frac{\overline{R}}{R_{\star}}=\left(\frac{\tau_0}{f}\right)^{1/(2n-\beta-s)}.
\end{equation}

It is easy to see from the expression that $\overline{R}$ is a function of both disk and stellar parameters. Besides, it is straightforward to derive the solution for an isothermal disk case, by simply setting $s=0$. A particularly interesting case is that of a reprocessing disk (optically thick, geometrically thin) at large radii, where $s=3/4$ \citep{lynden1974}. Realistic radiative transfer simulations indicate that such thermal description is approximately valid for the innermost region of the disk ({\it e.g.}, \citealt{carciofi2006}).

More explicitly, the effective radius depends on $\lambda$ and $\rho_0$ as

\begin{equation}
\overline{R}\propto \lambda^{(2+u)/(2n-\beta-s)}\times \rho_0^{2/(2n-\beta-s)},
\end{equation}
where we have used the definition

\begin{equation}
u\equiv\frac{d\ln(g+b)}{d\ln\lambda}.
\end{equation}

By choosing the parameters $n=3.5$, $\beta=1.5$ and $s=3/4$, and if we assume that $u\approx 0$, we have that $\overline{R}\propto (\lambda\times \rho_0)^{0.42}$. This means that the disk appears to be bigger either when observed at longer wavelengths or for higher disk densities.\\

Once the effective radius relation is determined, we can finally make use of the dual opacity regime assumed for the disk to obtain the following expression for the specific intensity

\begin{equation}
I_{\lambda}(\varpi)=\frac{2c\,k\,T_{\star}}{\lambda^4} \left\{ \begin{array}{ll}
1 & \textrm{$\varpi\leq R_{\star}$}\\
f\,(\varpi/R_{\star})^{-s} & \textrm{$R_{\star}<\varpi\leq\overline{R}$}\\
f\,(\overline{R}/R_{\star})^{-s}\,(\varpi/\overline{R})^{-2n+\beta} & \textrm{$\varpi>\overline{R}$}\\
\end{array} \right.
\end{equation}
where we again consider the Rayleigh-Jeans limit. For simplicity, the star was assumed to be spherical and of uniform brightness ({\it i.e.}, no limb darkening was considered). The flux expression can then be derived by the direct integration over the surface emitting area:

\begin{equation}
F_{\lambda} = F_{\lambda}^{\star} \left[ \frac{2-s-2f}{2-s} + 2f\frac{(2n-\beta-s)}{(2-s)(2n-\beta-2)}\left(\frac{\overline{R}}{R_{\star}}\right)^{2-s} \right],
\end{equation}
where the stellar flux is defined as

\begin{equation}\label{flux}
F_{\lambda}^{\star}=\frac{2\pi\,c\, k T_{\star}}{\lambda^4}\left(\frac{R_{\star}}{d} \right)^2.
\end{equation}

From this simple equation, it is straightforward to derive some other useful expressions, such as the IR excess $Z_{\nu}=F_{\lambda}/F_{\lambda}^{\star}$ and the spectral slope $\alpha_{IR}=-d\ln F_{\lambda}/d\ln\lambda$. A complete and more general derivation (extended to other inclinations, not restricted to the Rayleigh-Jeans approximation) of this set of equations will be presented by \citeauthor{vieira2015} (in preparation).

\section{Discussion}

\begin{figure}[!t]
\begin{center}
\includegraphics[width=1 \columnwidth,angle=0]{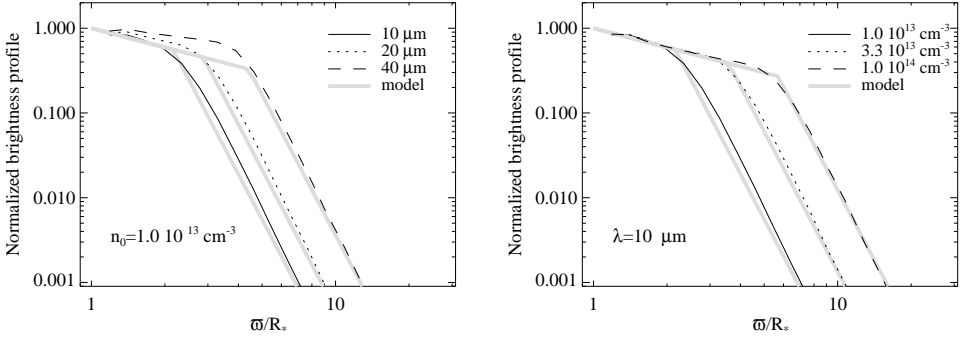}
\caption{Fit of brightness profiles computed with {\ttfamily HDUST} by analytical simple model, for different wavelengths and base densities. The symbols represent the values extracted from the simulation, while the full lines indicate the pseudo-photosphere model ($s=3/4$).}
\label{loci_fit}
\end{center}
\end{figure}

\begin{figure}[!t]
\begin{center}
\includegraphics[width=0.8 \columnwidth,angle=0]{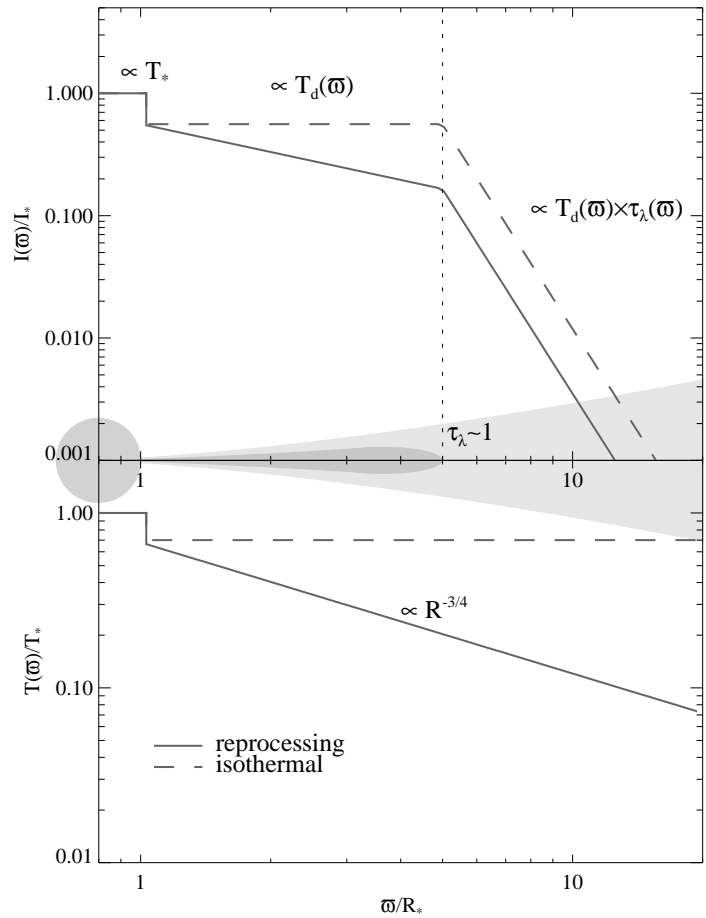}
\caption{Ilustration of direct relation between the brightness profile and temperature structure of the disk (in the Rayleigh-Jeans regime), for both isothermal and reprocessing disk temperature structures.}
\label{iconic}
\end{center}
\end{figure}

Figure~\ref{loci_fit} confronts the {\ttfamily HDUST} results to those obtained from the pseudo-photosphere model (for $s=3/4$). We can see that the analytical model  successfully reproduces the main properties of the numerical brightness profiles. Of course, the model is too simple to reproduce all the details from the realistic simulations, and there are some second order effects not taken into account that also modify the disk brightness. To cite a few limitations, the transition between the two regions is actually more smooth than the adopted one, and the temperature structure is much more complicated than a simple power-law ({\it e.g.}, \citealt{carciofi2006}, Fig.~8).\\

However, the general good agreement suggests that the adopted assumptions, albeit quite simple, are sufficiently realistic. The transition between the two components occurs very close to the position predicted by Eq.~\ref{ref}.\\

Besides demonstrating that a dual regime satisfactorily describes the disk intensity, our model could also show that the brightness profile is directly proportional to the disk thermal structure (in the Rayleigh-Jeans limit). This property is ilustrated in Fig.~\ref{iconic}, for both reprocessing and isothermal disk cases.\\

Finally, it is important to mention that the disk brightness profile itself determines both measured fluxes (Eq.~\ref{flux}) and interometric observations (obtained from the Fourier Transform of the brightness distribution). Consequently, the IR observables are directly affected by the {\it non-isothermal} disk structure. Our model quantifies the brightness-temperature connexion, and allows us to explore detectable signatures of the disk non-isothermality.

\section{Conclusions}

In the present work, we have introduced the pseudo-photosphere model. It consists of a disk dual brightness regime, with a characteristic effective radius separating them. By assuming a simplified power-law thermal structure, we were able to derive expressions for $\overline{R}$, the specific intensity and the emergent flux.

Although the disk thermal structure is much more complex than a simple power-law, the model specific intensities were shown to be very close to those obtained from realistic radiative transfer simulations.

In particular, we have derived the relation between the thermal structure and the disk brightness profile. Such relation allows us to study the effects of the disk non-isothermality on the observed fluxes and visibilities. A more general description of our analytical model and its applications will soon be presented in forthcoming papers.

\acknowledgements This work made use of the computing facilities of the Laboratory of Astroinformatics (IAG/USP, NAT/Unicsul), whose purchase was made possible by the Brazilian agency FAPESP (grant No 2009/54006-4) and the INCT-A.  R.~G.~V. akcnowledges the support from the organizers of the event, and from FAPESP (grant No 2012/20364-4). A.~C.~C acknowledges support from CNPq (grant No 307076/2012-1).

\bibliography{biblio}  

\begin{thebibliography}{}
\expandafter\ifx\csname natexlab\endcsname\relax\def\natexlab#1{#1}\fi
\expandafter\ifx\csname url\endcsname\relax
  \def\url#1{\texttt{#1}}\fi
\expandafter\ifx\csname urlprefix\endcsname\relax\def\urlprefix{URL }\fi
\providecommand{\eprint}[2][]{\url{#2}}

\bibitem[{{Allen}(1977)}]{allen1977}
{Allen}, K.~W. 1977, {Astrophysical quantities.}

\bibitem[{{Carciofi}(2011)}]{carciofi2011}
{Carciofi}, A.~C. 2011, in IAU Symposium, edited by C.~{Neiner}, G.~{Wade},
  G.~{Meynet}, \& G.~{Peters}, vol. 272 of IAU Symposium, 325.
  \eprint{1009.3969}

\bibitem[{{Carciofi} \& {Bjorkman}(2006)}]{carciofi2006}
{Carciofi}, A.~C., \& {Bjorkman}, J.~E. 2006, \apj, 639, 1081.
  \eprint{astro-ph/0511228}

\bibitem[{{Cot\'e} \& {Waters}(1987)}]{cote1987}
{Cot\'e}, J., \& {Waters}, L.~B.~F.~M. 1987, \aap, 176, 93

\bibitem[{{Gehrz} et~al.(1974){Gehrz}, {Hackwell}, \& {Jones}}]{gehrz1974}
{Gehrz}, R.~D., {Hackwell}, J.~A., \& {Jones}, T.~W. 1974, \apj, 191, 675

\bibitem[{{Lamers} \& {Waters}(1984)}]{lamers1984}
{Lamers}, H.~J.~G.~L.~M., \& {Waters}, L.~B.~F.~M. 1984, \aap, 136, 37

\bibitem[{{Lee} et~al.(1991){Lee}, {Osaki}, \& {Saio}}]{lee1991}
{Lee}, U., {Osaki}, Y., \& {Saio}, H. 1991, \mnras, 250, 432

\bibitem[{{Lynden-Bell} \& {Pringle}(1974)}]{lynden1974}
{Lynden-Bell}, D., \& {Pringle}, J.~E. 1974, \mnras, 168, 603

\bibitem[{{Rivinius} et~al.(2013){Rivinius}, {Carciofi}, \&
  {Martayan}}]{rivi2013}
{Rivinius}, T., {Carciofi}, A.~C., \& {Martayan}, C. 2013, \aapr, 21, 69.
  \eprint{1310.3962}

\bibitem[{{Struve}(1931)}]{struve1931}
{Struve}, O. 1931, \apj, 73, 94

\bibitem[{{Vieira} et~al.(prep){Vieira}, {Carciofi}, \&
  {Bjorkman}}]{vieira2015}
{Vieira}, R.~G., {Carciofi}, A.~C., \& {Bjorkman}, J.~E. in~prep.

\bibitem[{{Waters}(1986)}]{waters1986}
{Waters}, L.~B.~F.~M. 1986, \aap, 162, 121

\bibitem[{{Wright} \& {Barlow}(1975)}]{wright1975}
{Wright}, A.~E., \& {Barlow}, M.~J. 1975, \mnras, 170, 41

\end{thebibliography}

\end{document}